\documentclass[preprint2,10.5pt]{aastex}
\begin{document}

\title{\large{\rm{STRENGTHENING THE OPEN CLUSTER DISTANCE SCALE VIA VVV PHOTOMETRY\thanks{Based on observations taken within the ESO VISTA Public Survey VVV, Programme ID 179.B-2002.}}}}
\author{D. Majaess$^{1,}$\altaffilmark{*}, D.~Turner$^{1}$, C. Moni Bidin$^{2}$, D. Geisler$^{2}$, J. Borissova$^{7}$, D. Minniti$^{3,4,5}$, C. Bonatto$^{8}$  \\ W. Gieren$^{2}$, G. Carraro$^{9}$, R. Kurtev$^{7}$,  F. Mauro$^{2}$, A-N. Chen\'e$^{2,7}$, D. W. Forbes$^{10}$, P. Lucas$^{6}$ \\ I. D\'ek\'any$^{3}$, R. K. Saito$^{3}$, M. Soto$^{11}$} 
\affil{$^{1}$Saint Mary's University, Halifax, Nova Scotia, Canada}
\affil{$^{2}$Universidad de Concepci\'on, Concepci\'on, Chile}
\affil{$^{3}$Pontificia Universidad Catolica de Chile, Santiago, Chile} 
\affil{$^{4}$Vatican Observatory, Vatican City, Italy}
\affil{$^{5}$Princeton University, Princeton, New Jersey, USA}
\affil{$^{6}$University of Hertfordshire, Hatfield, United Kingdom}
\affil{$^{7}$Universidad de Valpara\'iso, Valpara\'iso, Chile}
\affil{$^{8}$Universidade Federal do Rio Grande do Sul, Porto Alegre, RS, Brazil.}
\affil{$^{9}$ESO, Santiago de Chile, Chile.}
\affil{$^{10}$Sir Wilfred Grenfell College, Memorial University, Corner Brook, Newfoundland, Canada.}
\affil{$^{11}$Universidad de la Serena, la Serena, Chile.}
\altaffiltext{*}{dmajaess@cygnus.smu.ca}

\begin{abstract}
Approximately $14$\% of known Galactic open clusters possess absolute errors $\le 20$\% as evaluated from $n\ge3$ independent distance estimates, and the statistics for age estimates are markedly worse.  That impedes such diverse efforts as calibrating standard candles and constraining masses for substellar companions. New data from the VVV survey may be employed to establish precise cluster distances with comparatively reduced uncertainties ($\le10$\%).  This is illustrated by deriving parameters for Pismis~19 and NGC~4349, two pertinent open clusters which hitherto feature sizable uncertainties ($60$\%). Fundamental parameters determined for Pismis~19 from new VVV $JHK_s$ photometry are $d=2.40\pm0.15$ kpc, $<{\it E_{J-H}}>=0.34\pm0.04$, and $\log{\tau}=9.05\pm0.10$, whereas for NGC~4349 the analysis yielded $d=1.63\pm0.13$ kpc, ${\it E_{J-H}}=0.09\pm0.02$, $\log{\tau}=8.55\pm0.10$.  The results exhibit a significant ($\ge 5 \times$) reduction in uncertainties, and indicate that: $i$) existing parameters for the substellar object NGC~4349 127b require revision, in part because the new cluster parameters imply that the host is $20$\% less-massive (${\cal M_{*}/M_{\sun}} \sim 3.1$);  $ii$) R Cru is not a member of NGC~4349 and should be excluded from period-Wesenheit calibrations that anchor the distance scale; $iii$) and results for Pismis~19 underscore the advantages gleaned from employing deep VVV $JHK_s$ data to examine obscured ($A_V \sim 4$) and differentially reddened intermediate-age clusters.
\end{abstract}

\section{{\rm \footnotesize INTRODUCTION}}
Approximately $30$\% of the 395 open clusters featuring $n\ge3$ independent distance estimates exhibit absolute errors $\ge 20$\%  \citep[][their Fig.~2]{pa06}.  There are $\ge 2 \times 10^3$ cataloged Galactic open clusters \citep{di02}, implying that merely $\sim 14$\% of the known sample possess errors $\le 20$\% as evaluated from three distance estimates.  The uncertainties permeate into analyses which rely on the cluster zero-point, such as the calibration of any constituent standard candles or substellar companions \citep{lm07,ma11b}.   Consider that published parameters for NGC~4349 span $d=900-2200$ pc and $\tau=0.1-0.6$ Gyr (\S \ref{s-ngc4349}).  Yet physical parameters for the substellar companion to TYC 8975-2601-1 \citep{lm07,ka08} rely on those inferred for the host from cluster membership (NGC~4349).  Furthermore, the nearer distance and younger age for NGC 4349 potentially imply cluster membership for the classical Cepheid R Cru, which lies within the cluster's corona. Establishing cluster membership would enable the subsequent calibration of Cepheid period-luminosity and period-Wesenheit relations \citep{tu10}.  Such functions bolster efforts to establish extragalactic distances and zero-point the SNe Ia scale \citep[e.g.,][]{pg04}.  The aforementioned examples underscore the broad ramifications of an uncertain cluster scale.  Admittedly, age estimates for open clusters are less reliable since a third exhibit absolute errors $>50$\% \citep[][$n\ge3$]{pa06}, and presumably the statistics worsen for obscured clusters near the Galatic plane.  

In this study, new VVV (VISTA Variables in the V\'{i}a L\'actea) $JHK_s$ photometry is employed to illustrate the marked improvement that can be achieved \textit{vis \`a vis} open cluster distances.  Two important clusters featuring particularly discrepant published parameters are examined, namely Pismis~19 and NGC~4349.  Distances for the clusters display a $\sim 60$\% spread and individual uncertainties of $\sim 30$\%.  Efforts to secure precise parameters for Pismis~19 via optical photometry have been complicated by differential reddening and $A_V\sim4$ magnitudes of obscuring dust. Parameters for Pismis~19 and NGC~4349 derived here exhibit a marked ($>5 \times$) reduction in uncertainties (\S \ref{s-analysis}), and highlight the advantages of using VVV data to determine reliable cluster distances and compliment existing efforts.

\section{{\rm \footnotesize VVV PHOTOMETRY}}
\begin{figure}[!t]
\begin{center}
\includegraphics[width=6cm]{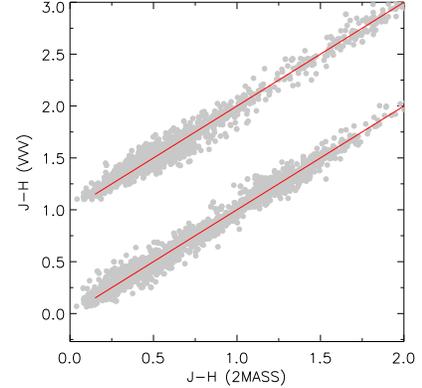} 
\caption{\small{A comparison between 2MASS and VVV $J-H$ photometry for the region encompassing Pismis~19 and NGC~4349.  A 1:1 correlation (red line) exists to within the uncertainties.  The data for Pismis~19 were deliberately offset from zero for presentation purposes.  Pertinent details regarding the pipeline employed here to process the VVV data are described in \citet{mb11} and Mauro et al. (2011, in prep.).}}
\label{fig-fov}
\end{center}
\end{figure}

The VVV survey aims to establish precise multi-epoch $JHK_s$ photometry for fields in the Galactic bulge and near the Galactic plane \citep[$\ell=295-10 \degr$,][]{mi10,cat11,sa11}.  VVV images exhibit increased angular resolution relative to 2MASS, and extend $\sim 4$ magnitudes fainter for Galactic disk stars.  The deep $JHK_s$ photometry facilitates isochrone fitting by revealing the target cluster's evolutionary morphology, which is particularly important when investigating highly reddened clusters.  The VVV survey will provide standardized (2MASS) $JHK_s$ photometry for stars in $\ge 3 \times 10^2$ open clusters and $\ge39$ globular clusters (e.g., M28). Details of the pipeline constructed to process and extract the VVV photometry employed here are discussed in Mauro et al. (2011, in prep.).  PSF photometry was performed using DAOPHOT and subsequently tied to 2MASS {\it JHK}$_s$ standards \citep[Fig.~\ref{fig-fov}, see also][]{mb11}.  However, as with any nascent large scale survey adjustments to the zero-point may occur as improvements and systematic errors are identified.

\subsection{{\rm \footnotesize THE ADVANTAGES OF $JHK_s$ PHOTOMETRY}}
Precise $JHK_s$ observations of stellar clusters are desirable since total and differential reddening are less deleterious than in the optical ($A_J\sim 0.2\times A_V$).  Sizable extinction may shift a significant fraction of the main-sequence near/beyond the limiting magnitude where uncertainties are largest.  Consider that merely $\sim 3^{\rm m}$ of the cluster sequence for Pismis 19 was sampled in existing optical surveys owing to significant reddening, which subsequently complicated efforts to establish the cluster zero-point.   The $JHK_s$ reddening vector provides viable solutions for the intrinsic colors of stars across much of the main-sequence.  The $JHK_s$ reddening vector can be determined from red clump stars along the line of sight \citep{sl08,ma11b}, and the ratio of total-to-selective extinction can be inferred in certain instances using red clump stars via the variable extinction method \citep[e.g.,][]{ma11b}.  \citet{sl08} demonstrated that $E(J-H)/E(H-K_s)$ is (rather) constant for dust occupying the inner Galaxy.  A consensus exists that any variations in the infrared would be marginal relative to that expected for the optical.

$JHK_s$ photometry is particularly suited for detecting and characterizing the heavily obscured pre main-sequence population of young clusters \citep{bb10}.  For later-type stars, $JHK_s$ photometry is relatively insensitive to variations in chemical composition \citep[e.g., the Hyades anomaly,][see also \citealt{sl09}]{tu79,ma11}.  That claim is supported in part by the establishment of seven benchmark open clusters ($d<250$ pc) which exhibit matching $JHK_s$ ZAMS and revised Hipparcos distances \citep[the Hyades, $\alpha$ Per, Praesepe, Coma Ber, IC 2391, IC 2609, and NGC 2451,][]{vl09,ma11}.  The zero-point of the Padova isochrone employed (\S \ref{s-analysis}) matches that scale, to within the uncertainties.  Isochrones, models, and the distance scale should be anchored (\& evaluated) using clusters where consensus exists, rather than the discrepant case (i.e. the Pleiades).  The 2MASS survey provides invaluable all-sky $JHK_s$ photometric standards.  A similar survey tied to Johnson-Cousins $UBVRI$ photometry is desirable.  $U$-band photometry is particularly challenging to standardize and zero-point errors are common \citep[\S \ref{s-ngc4349}, see also][]{cc01}.  However, $UBV$ color-color analyses permit crucial dereddening for younger stars.

In summary, the VVV survey is aptly tailored to foster cluster research \citep{mi11,bo11,mb11,ma11b}. Admittedly, acquiring precise and standardized $UBVJHK_s$ photometry is ideal, and enables the characterization of potential systemic errors. $UBV$ data by Turner/Forbes (unpublished) and \citet[][in press]{ca11} are employed to corroborate parameters determined via the VVV photometry.  

\section{{\rm \footnotesize ANALYSIS}}
\label{s-analysis}
\subsection{{\rm \footnotesize PISMIS~19}}
\label{s-pismis19}
Pismis~19 (Fig.~\ref{fig-fov}) is a heavily reddened open cluster \citep{pi98,cm04}.  The cluster's non-symmetric appearance in optical images is indicative of differential reddening.  \citet{pi98} and \citet{cm04} acquired $BVI$ photometry  for Pismis~19 stars.  However, separate conclusions were reached regarding the cluster's fundamental parameters.  \citet{pi98} determined the following: $E(B-V)=1.45\pm0.10$, $d=2.40\pm0.88$ kpc, $\tau=1.0\pm0.2$ Gyr, whereas \citet{cm04} obtained $E(B-V)=1.48\pm0.15$, $d=1.5\pm0.4$,  $\tau\simeq0.8$ Gyr.  The reddenings and distances agree within the mutual uncertainties, however, the individual uncertainties are large owing to the challenging task of analyzing highly obscured clusters solely via optical photometry.  \citet{ca11} built upon \citet{pi98} and \citet{cm04} analyses by obtaining deeper photometry, and derived $d=2.5\pm0.5$ kpc. 

\begin{figure}[!t]
\begin{center}
\includegraphics[width=6.8cm]{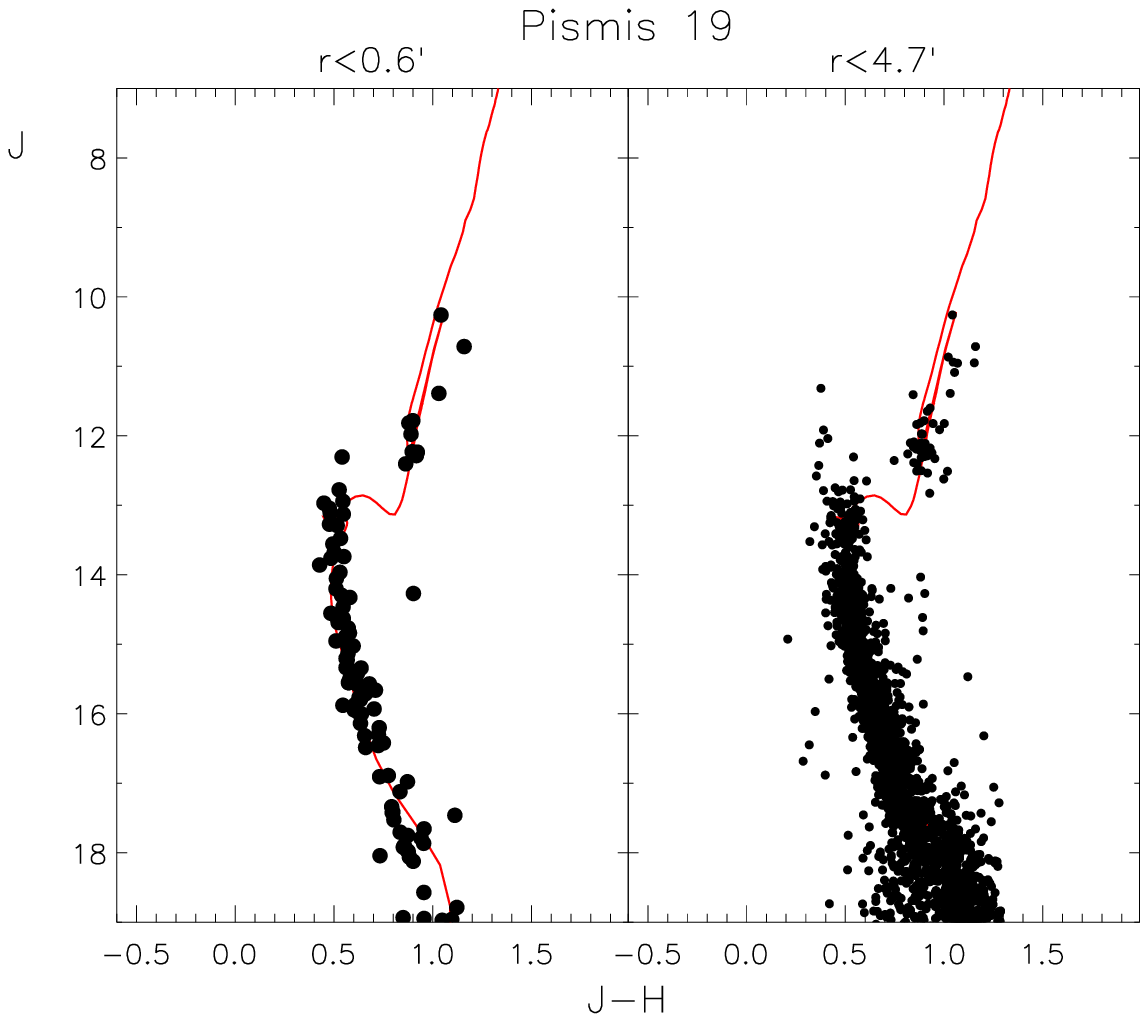} 
\includegraphics[width=5.1cm]{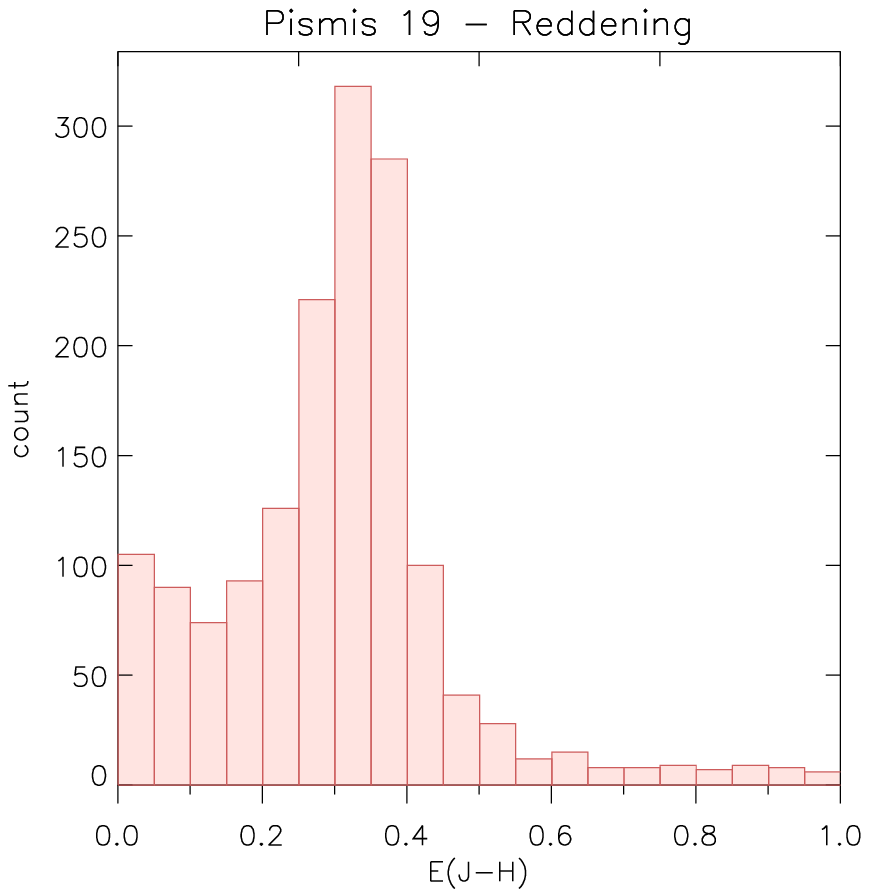} 
\caption{\small{VVV color-magnitude diagrams for Pismis~19 at varying radii. $JHK_s$ Padova isochrones \citep{bo04} were employed.  Individual reddenings were determined via the approach outlined in \S \ref{s-pismis19}.  Photometric errors and binary contamination artificially broaden the breadth of the differential reddening.}}
\label{fig-dr}
\end{center}
\end{figure}

Individual reddenings for stars in Pismis~19 were determined as follows. Any point on the dereddening line (dl) for the $i^{th}$ star is given by:$
(J-H)_i=E_{J-H}/E_{H-K_s} \times (H-K_s)_i + b$; $b=(J-H)_i-  E_{J-H}/E_{H-K_s} \times (H-K_s)_i$; $(J-H)_{dl}=E_{J-H}/E_{H-K_s} \times (H-K_s)_{dl} + b$.  The intersect between the dereddening line and the intrinsic relation was determined by minimizing the difference as a function of $(H-K_s)_0$:$
|(J-H)_{dl}-(J-H)_{0,z}| =E_{J-H}/E_{H-K_s} \times (H-K_s)_{0,z} 
+(J-H)_i -E_{J-H}/E_{H-K_s} \times (H-K_s)_i -(J-H)_{0,z}$.The reddening vector ($E_{J-H}/E_{H-K_s}$) characterizing dust along the line of sight was derived by tracking deviations of red clump stars from their mean intrinsic color owing to extinction.  The mean intrinsic color was adopted from \citet{ma11b}, who inferred the result from nearby red clump stars ($d\le50$ pc) with revised Hipparcos parallaxes \citep{vl07}.  The reddening vector determined from red clump stars is $E(J-H)/E(H-K_s)=2.02$.  That result agrees with a determination for the region from 2MASS photometry \citep{sl08}. A reddening diagram was subsequently compiled (Fig.~\ref{fig-dr}), and the mean reddening is $<E(J-H)>=0.34\pm0.04$.   The cluster stars terminate near F2 according to the intrinsic {\it JHK}$_s$ relation of \citet{sl09}.

A color-magnitude diagram was compiled for Pismis~19 stars surrounding J2000 coordinates of 14:30:40.54 -60:53:32.2 (Fig.~\ref{fig-dr}).\footnote{The coordinates cited for the cluster center in simbad require updating.}  A $\log{\tau}=9.05\pm0.10$ Padova isochrone \citep{bo04} was adopted based on the reddening and spectral types inferred (Fig.~\ref{fig-dr}), and since that age provides an evolutionary track which matches cluster members ranging from $\sim$M0 dwarfs to evolved stars. A precise fit was obtained owing to several factors.  First, two of three free parameters associated with isochrone fitting were constrained by the color-color analysis, namely the reddening and age (spectral type at the turnoff).  The remaining parameter (excluding metallicity) is the shift required in magnitude space to overlay the isochrone upon the data.  The best fit and uncertainties were established via the traditional visual approach \citep[e.g.,][]{cm04,bb10}, and the latter represents the limit where a mismatch is clearly perceived.  \citet[][and references therein]{pa06} note that errors tied to isochrone fitting via computer algorithms are comparable to those associated with the traditional approach.  Secondly, field star contamination was mitigated since the surface density of cluster members is an order of magnitude larger, and furthermore, the cluster members occupy a heavily reddened locus separated from less-reddened field stars (Fig.~\ref{fig-dr}).  Lastly, the deep VVV photometry provided excellent anchor points for isochrone fitting. 

The final parameters for Pismis~19 are: $d=2.40\pm0.15$ kpc, $<{\it E(J-H)}>=0.34\pm0.04$, and $\log{\tau}=9.05\pm0.10$.  The distance is tied to a ratio of total to selective extinction ({\it R}) derived by \citet{ma11} \citep[see also][and references therein]{bo04}.  The distance derived here agrees with the latest estimate from optical photometry \citep{ca11}.  

\subsection{{\rm \footnotesize NGC~4349}}
\label{s-ngc4349}

\citet{ko56} and \citet{kr57} noted that the $5.8^{\rm d}$ classical Cepheid R Cru may be a member of NGC~4349.  That assessment was based in part on the Cepheid's proximity and brightness relative to cluster members.  Cepheids are typically among the foremost evolved members of their host clusters.  \citet{lo61} employed $UBV$ photographic photometry to derive cluster parameters of: $d=1700$ pc and $\tau=600$ Myr.  \citet{fe63} obtained photoelectric $BV$ photometry from the Cape Observatory and established a cluster distance of $d=900$ pc, for $E(B-V)=0.31$.  The distance to NGC~4349 cited by \citet{fe63} is approximately half that derived by \citet{lo61}. \citet{fe63} concluded that R Cru is unassociated with NGC~4349 since the Cepheid lies towards the cluster's periphery.  Cluster Cepheids known during that era had been discovered near the cluster center (e.g., S Nor/NGC 6087).  Incidentally, the distance to NGC~4349 established by \citet{fe63} is consistent with that inferred for R Cru from present day period-Wesenheit relations \citep{be07,ma11b}.  \citet{li68} revised the \citet{lo61} age for NGC~4349 downward to $\log{\tau}=8.04$. \citet{lm94} computed the following properties for NGC~4349 based on a reanalysis of existing photometry: $d=2176$ pc, $E(B-V)=0.384$, and $\log{\tau}=8.315$.  In sum, published parameters for NGC~4349 span $d=900-2200$ pc.

\begin{figure}[!t]
\begin{center}
\includegraphics[width=6.9cm]{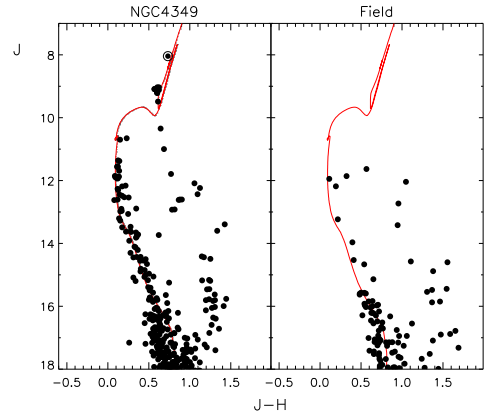} 
\caption{\small{Color-magnitude diagrams constructed for NGC~4349 and an adjacent comparison field using VVV/2MASS $JHK_s$ photometry. The circled dot near the tip of the giant branch is TYC 8975-2606-1, which hosts a substellar companion \citep{lm07}. To mitigate contamination the CMDs feature stars within $r<1.2 \arcmin$.  Seven evolved red stars beyond that radius were added to the CMD for NGC~4349.}}
\label{fig-cm}
\end{center}
\end{figure}

 A reddening vector of $E(J-H)/E(H-K_s)=2.04$ was determined from red clump stars along the line of sight \citep[see also][]{sl08}.  The reddening vector was subsequently adopted to establish a reddening of {\it E(J-H)}$=0.09\pm0.02$.  Stars catalogued by \citet{lo61} as likely cluster members were employed to derive that result.   New photoelectric $UBV$ photometry\footnote{Obtained with the 0.6 m Helen Sawyer Hogg Telescope which was stationed at Cerro Las Campanas, Chile.} obtained for stars in NGC 4349 were likewise used to constrain the cluster reddening, and age.  A comparison between that photoelectric $UBV$ photometry and the photographic photometry of \citet{lo61} reveals the latter is offset from the standard system: $B-V=(1.02\pm0.02) \times (B-V)_{{\rm L61}}-0.02\pm0.02$; 
$U-B=(0.96\pm0.02) \times (U-B)_{{\rm L61}}+0.09\pm0.01$; $V=(-0.015\pm0.03) \times (B-V)_{{\rm L61}}+0.06\pm0.02+V_{{\rm L61}}$.
The offset may partly explain the difference between the distances inferred from the $UBV$ photometry of \citet{lo61} and the present analysis.  Applying an intrinsic  $UBV$ color-color relation to the corrected data yields a reddening of $E(B-V)=0.32\pm0.03$.  The canonical extinction law was employed, and may be refined once spectroscopic observations are available.  Stars in NGC~4349 terminate near B8-A0 according to the intrinsic {\it JHK}$_s$ and $UBV$ relations \citep[e.g.,][]{sl09,tu11}.  Published reddenings for R Cru \citep{fe90} are nearly half that derived for the cluster, implying that the Cepheid lies in the foreground. That is consistent with the Cepheid's parameters as inferred from the latest period-Wesenheit relations \citep[e.g.,][]{be07}, which indicate that R Cru is less than 1 kpc distant.  

A color-magnitude diagram for NGC~4349 is shown in Fig.~\ref{fig-cm}.  A $\log{\tau}=8.55\pm0.10$ Padova isochrone \citep{bo04} was adopted based on the reddening and spectral types inferred from the color-color diagram, and since that age provides an evolutionary track which aptly matches both bluer and redder evolved members.  NGC~4349 features evolved stars brighter than the saturation limit of the VVV survey.  Therefore, $JHK_s$ photometry for these stars were taken from 2MASS.  The color-magnitude diagram for NGC 4349 was restricted to stars within $\le 1.2 \arcmin$ to mitigate field contamination.  The final parameters for NGC~4349 are: $d=1.63\pm0.13$ kpc, ${\it E(J-H)}=0.09\pm0.02$, and $\log{\tau}=8.55\pm0.10$.  A ratio of total to selective extinction ({\it R}) was adopted from \citet{ma11b} \citep[see also][and references therein]{bo04}.   The distance and reddening agree with that established by \citet[][$d=1.74\pm0.65$ kpc and $E_{B-V}=0.34\pm0.03$]{cl89}.

In their comprehensive survey \citet{lm07} discovered that TYC 8975-2606-1 hosts a substellar companion (designated NGC~4349 127b).  \citet{lm07} adopted cluster parameters of $d=2200$ pc and $\tau=200$ Myr, which implied a $3.9{\cal M_{\sun}}$ host.  However, the distance and age established here for NGC~4349 are $30$\% nearer and $150$ Myr older, respectively.  The parent star exhibits the following parameters according to the Padova isochrone applied: ${\cal M_{*}/M_{\sun}}\sim3.1$ and $\log {L/L_{\sun}}\sim2.7$.  Yet the principal source of uncertainty hindering a reliable determination of the orbital parameters remains the sparsely sampled radial velocity curve \citep{lm07}, as indicated by simulations conducted using the \textit{Systemic Console} \citep{me09}.   Nevertheless, a minimum mass for the substellar companion of ${\cal M/M_{J}}\sim 17$ was obtained.  \citet{ka08} derived an X-ray luminosity for the system in order to evaluate whether giant planets in close proximity to the host are catalysts for magnetic activity.  That determination was based on a distance to NGC~4349 of $d=2176$ pc, thereby reaffirming the importance of a precise distance scale.  

\section{{\rm \footnotesize CONCLUSION}}
VVV $JHK_s$ observations may be employed to help establish precise cluster distances that feature comparatively reduced uncertainties ($\le 10$\%).  That is illustrated by deriving fundamental parameters for Pismis~19 and NGC~4349, two important clusters which hitherto exhibit sizable uncertainties ($60$\%, \S \ref{s-pismis19} and \ref{s-ngc4349}).  A precise distance determination for Pismis~19 from optical photometry was hampered in part by significant reddening (Fig.~\ref{fig-dr}, $A_V \sim 4$).  The existing ambiguity surrounding the distance to NGC~4349 ensured that the pertinence of invaluable putative constituents were mitigated (i.e., the classical Cepheid R Cru and a substellar companion for the  member TYC 8975-2601-1).  Parameters derived for Pismis~19 are: $d=2.40\pm0.15$ kpc, $<{\it E(J-H)}>=0.34\pm0.04$, $\log{\tau}=9.05\pm0.10$ (Fig.~\ref{fig-dr}), whereas NGC~4349 exhibits $d=1.63\pm0.10$ kpc, ${\it E(J-H)}=0.09\pm0.02$, $\log{\tau}=8.55\pm0.10$ (Fig.~\ref{fig-cm}).  The nature of the VVV survey ensured that the revised results, which have pertinent ramifications, compliment existing estimates and display a marked improvement ($\ge 5 \times$) in precision. New VVV $JHK_s$ for stars in NGC~4349 and Pismis~19 imply that: existing physical parameters derived for NGC~4349 127b need to be redetermined in part since the mass for the host star was revised downward to ${\cal M_{*}/M_{\sun}}\sim 3.1$ (\S \ref{s-ngc4349}); the classical Cepheid R Cru is not a member of NGC~4349 (\S \ref{s-ngc4349});  and VVV $JHK_s$ photometry is particularly suited for constraining parameters of obscured and differentially reddened intermediate-age clusters (e.g., Pismis~19, $A_V \sim 4$, Fig.~\ref{fig-dr}).   
 
The VVV and UKIDSS surveys \citep{lu08,mi10} may be employed to achieve significant gains toward strengthening the open cluster distance scale.  Yet considerable work remains, and unknown systematic errors may be discovered given the nascent nature of the aforementioned surveys.  Consequently, obtaining independent multiband observations are desirable to corroborate derived cluster parameters \citep[e.g.,][]{ca11}.  

\subsection*{{\rm \scriptsize ACKNOWLEDGEMENTS}}
\scriptsize{DM is grateful to the following individuals and consortia whose efforts lie at the foundation of the research: 2MASS, P. Stetson (DAOPHOT), W. Lohmann, WEBDA (E. Paunzen), DAML (W. Dias), C. Lovis, M. Mayor, CDS, arXiv, and NASA ADS.  We gratefully acknowledge use of data from the ESO Public Survey programme ID 179.B-2002 taken with the VISTA telescope, the Cambridge Astronomical Survey Unit, and funding from the FONDAP Center for Astrophysics 15010003, the BASAL CATA Center for Astrophysics and Associated Technologies PFB-06, the MILENIO Milky Way Millennium Nucleus from the Ministry of Economics ICM grant P07-021-F, and Proyecto FONDECYT Regular 1090213. WG, CMB, and DG are grateful for support from the Chilean Center for Astrophysics FONDAP 15010003 and the BASAL Centro de Astrofisica y Tecnologias Afines (CATA) PFB-06/2007. RK acknowledges support from Proyecto DIUV23/2009, Universidad de Valparaiso.  RS acknowledges financial support from CONICYT through GEMINI Project Nr. 32080016.}


\begin{thebibliography}{}\setlength{\itemsep}{-1.6mm}
\bibitem[Benedict et al.(2007)]{be07} Benedict, G.~F., 
McArthur, B.~E., Feast, M.~W., et al.\ 2007, \aj, 133, 1810
\bibitem[Bonatto et al.(2004)]{bo04} Bonatto, C., Bica, E., \& Girardi, L.\ 2004, \aap, 415, 571 
\bibitem[Bonatto \& Bica(2010)]{bb10} Bonatto, C., \& Bica, E.\ 2010, \aap, 516, A81 
\bibitem[Borissova et al.(2011)]{bo11} Borissova, J., et al.\ 2011, \aap, 532, A131 
\bibitem[Carraro \& Munari(2004)]{cm04} Carraro, G., \& Munari, U.\ 2004, \mnras, 347, 625 
\bibitem[Carraro(2011)]{ca11} Carraro, G.\ 2011, A\&A, in press
(arXiv:1111.0579) 
\bibitem[Catelan et al.(2011)]{cat11} Catelan, M., Minniti, 
D., Lucas, P.~W., et al.\ 2011, RR Lyrae Stars, Metal-Poor Stars, and the 
Galaxy, 145 
\bibitem[Claria \& Lapasset(1989)]{cl89} Claria, J.~J., \& Lapasset, E.\ 1989, \mnras, 241, 301 
\bibitem[Cousins \& Caldwell(2001)]{cc01} Cousins, A.~W.~J., \& Caldwell, J.~A.~R.\ 2001, \mnras, 323, 380 
\bibitem[Dias et 
al.(2002)]{di02} Dias, W.~S., Alessi, B.~S., Moitinho, A., \& L{\'e}pine, J.~R.~D.\ 2002, \aap, 389, 871 
\bibitem[Fernie(1963)]{fe63} Fernie, J.~D.\ 1963, The 
Observatory, 83, 33 
\bibitem[Fernie(1990)]{fe90} Fernie, J.~D.\ 1990, \apjs, 72, 
153
\bibitem[Lindoff(1968)]{li68} Lindoff, U.\ 1968, Arkiv for Astronomi, 5, 1 
\bibitem[Lohmann(1961)]{lo61} Lohmann, W.\ 1961, 
Astronomische Nachrichten, 286, 105 
\bibitem[Loktin \& Matkin(1994)]{lm94} Loktin, A.~V., \& Matkin, N.~V.\ 1994, A\&AT, 4, 153 
\bibitem[Lovis 
\& Mayor(2007)]{lm07} Lovis, C., \& Mayor, M.\ 2007, \aap, 472, 657 
\bibitem[Lucas et al.(2008)]{lu08} Lucas, P.~W., et al.\ 2008, \mnras, 391, 136 
\bibitem[Kashyap et al.(2008)]{ka08} Kashyap, V.~L., Drake, 
J.~J., \& Saar, S.~H.\ 2008, \apj, 687, 1339 
\bibitem[Kholopov(1956)]{ko56} Kholopov, P.~N.\ 1956, 
Peremennye Zvezdy, 11, 325 
\bibitem[Kraft(1957)]{kr57} Kraft, R.~P.\ 1957, \apj, 126, 
225 
\bibitem[Majaess et al.(2011a)]{ma11} Majaess, D., Turner, 
D., Lane, D., \& Krajci, T.\ 2011 (a), JAAVSO, 39, 219 
\bibitem[Majaess et al.(2011b)]{ma11b} Majaess, D., Turner, 
D., Moni Bidin, C., et al.\ 2011 (b), \apjl, 741, L27 
\bibitem[Meschiari et al.(2009)]{me09} Meschiari, S., Wolf, 
A.~S., Rivera, E., et al.\ 2009, \pasp, 121, 1016 
\bibitem[Minniti et al.(2010)]{mi10} Minniti, D., et al.\ 2010, New Astronomy, 15, 433 
\bibitem[Minniti et al.(2011)]{mi11} Minniti, D., et al.\ 2011, \aap, 527, A81 
\bibitem[Moni Bidin et 
al.(2011)]{mb11} Moni Bidin, C., Mauro, F., Geisler, D., et al.\ 2011, \aap, 535, A33 
\bibitem[Paunzen 
\& Netopil(2006)]{pa06} Paunzen, E., \& Netopil, M.\ 2006, \mnras, 371, 1641 
\bibitem[Piatti et al.(1998)]{pi98} Piatti, A.~E., 
Clari{\'a}, J.~J., Bica, E., Geisler, D., 
\& Minniti, D.\ 1998, \aj, 116, 801 
\bibitem[Pietrzy{\'n}ski \& Gieren(2004)]{pg04} Pietrzy{\'n}ski, G., \& Gieren, W.\ 2004, IAU Colloq.~193: Variable Stars in the Local Group, 310, 87 
\bibitem[Saito et al.(2011)]{sa11} Saito, R.~K., Hempel, M., 
Minniti, D., et al.\ 2011, A\&A, in press (arXiv:1111.5511)
\bibitem[Strai{\v z}ys 
\& Laugalys(2008)]{sl08} Strai{\v z}ys, V., \& Laugalys, V.\ 2008, Baltic Astronomy, 17, 253 
 \bibitem[Strai{\v z}ys \& Lazauskait{\.e}(2009)]{sl09} Strai{\v z}ys, V., \& Lazauskait{\.e}, R.\ 2009, Baltic Astronomy, 18, 19 
\bibitem[Turner(1979)]{tu79} Turner, D.~G.\ 1979, \pasp, 91, 
642 
\bibitem[Turner(2010)]{tu10} Turner, D.~G.\ 2010, \apss, 326, 219 
\bibitem[Turner(2011)]{tu11} Turner, D.~G.\ 2011, RMxAA, 
47, 127
\bibitem[van Leeuwen(2007)]{vl07} van Leeuwen, F.\ 2007, \aap, 474, 653 
\bibitem[van Leeuwen(2009)]{vl09} van Leeuwen, F.\ 2009, \aap, 497, 209 
\end{thebibliography}
\end{document}